\documentclass[aps,prl,twocolumn, groupedaddress]{revtex4-1}
\usepackage{color,soul}
\usepackage{amssymb}
\usepackage{graphicx}
\usepackage{bm}        
\usepackage{amssymb}   
\usepackage{amsfonts}
\usepackage{amsmath,mathtools} 
\usepackage{breqn} 
\usepackage{multirow}
\usepackage{array}
\usepackage{booktabs}
\usepackage{textcomp}
\usepackage{csquotes}

\begin {document}


\title{\large \bf Formation of  tungsten carbide by Focussed Ion Beam process : A route to high magnetic field resilient patterned superconducting nanostructures}
\author{Himadri Chakraborti$^1$, Bhanu P. Joshi$^{1,2}$, Chanchal K. Barman$^1$, Aditya K. Jain$^{1,3}$, Buddhadeb Pal$^{1,4}$, Bikash C. Barik$^1$, Tanmay Maiti$^5$, R\"{u}diger Schott$^6$, Andreas D. Wieck$^6$, M.J.N.V. Prasad$^7$, S. Dhar$^{1}$,  Hridis K. Pal$^1$, Aftab Alam$^{1}$ }
\author{K$.$ Das Gupta$^{1}$}
\email{kdasgupta@phy.iitb.ac.in}
\affiliation{$^{1}$Department of Physics$,$ Indian Institute of Technology Bombay$,$ Mumbai $-$ 400076$,$ India}

\affiliation{$^{2}$International Research Center MagTop$,$ Institute of Physics$,$ Polish Academy of Sciences$,~$ Al. Lotnik{\'o}w$~$ 32/46$,$ PL$-$02$-$668 Warszawa$,$ Poland}
\affiliation{$^{3}$Department of Physics$,$ Royal Holloway $,$ University of London$,$ Surrey $-$ TW20 0EX$,$ United Kingdom}

\affiliation{$^{4}$S. N. Bose National Centre for Basic Sciences$,$Kolkata$,$ West Bengal $-$700106$,$ India}

\affiliation{$^{5}$Saha Institute of Nuclear Physics$,$ HBNI$,$ 1/AF Bidhannagar$,$ Kolkata $-$ 700064$,$ India}

\affiliation{$^{6}$Lehrstuhl f\"{u}r Angewandte Festk\"{o}rperphysik $,$ Ruhr Universit\"{a}t Bochum $,$  D-44801 Bochum $,$  Germany}

\affiliation{$^{7}$Department of Metallurgical Engineering and Materials
Science$,$ Indian Institute of Technology Bombay$,$ Mumbai $-$ 400076$,$
India}



\begin{abstract}
	
A scale for magnetic field resilience of a superconductor is set by the  paramagnetic limit. Comparing the condensation energy of the Bardeen-Cooper-Schrieffer (BCS) singlet ground state with the  paramagnetically polarised state suggests that for an applied field
${\mu_0}H > 1.8~T_c$ (in SI),  singlet pairing is not  energetically favourable.  Materials exceeding or approaching this limit are interesting from fundamental and technological perspectives. This may be a potential indicator  of triplet superconductivity, Fulde-Ferrell-Larkin-Ovchinnikov (FFLO) pairing and other mechanisms involving  topological aspects of surface states, and also allow Cooper pair injection at high magnetic fields. We have analysed the microscopic composition of such a material arising from an unexpected source. A microjet of an organo-metallic gas, $\rm {W[(CO)_6]}$ can be decomposed by  gallium ion-beam, leaving behind a track of  complex residue of gallium, tungsten and carbon with
remarkable superconducting properties, like an upper critical field, $H_{c2} > 10~{\rm T} $, above its paramagnetic limit. We  carried out Atomic probe tomography to  establish the formation of nano-crystalline tungsten carbide (WC)  in the tracks and the  absence of free tungsten.  Supporting calculations show for  Ga  distributed on the surface of WC, its s,p-orbitals enhance the density of states near the Fermi energy. The observed variation of $H_{c2}(T)$ does not show features typical of enhancement of critical field due to granularity. Our observations may be significant in the context of some recent theoretical calculation of the band structure of WC and experimental observation of superconductivity in WC-metal interface.


\end{abstract}


\maketitle

A superconducting state that survives at high magnetic fields opens up interesting possibilities - {\it e.g.} potential coexistance of quantum Hall edge states and Cooper-pairs \cite{lee2017inducing, yacoby2017inducing}. However  superconductivity  is suppressed by a magnetic field in two ways. The kinetic energy of the shielding currents builds up with increasing external field and acts against the condensation energy. Secondly the magnetic field tries to align the spins and  making the spin-singlet state unfavourable. Compared to the free electron Fermi sea, the paramagnetically polarised state is lower in energy by $\dfrac{{\mu_B}^{2}D(E_F)\left( \mu_0{H}\right)^2}{2}$, where $\mu_B$ is the Bohr magneton and $D(E_F)$ is the density of states at the Fermi level. The BCS singlet ground state has a condensation energy $\dfrac{D(E_F)\Delta(0)^2}{2}$,  where $\Delta(0)=1.76{k_B}{T_c}$ is the BCS  gap. $T_c$ is the critical temperature. For ${\mu_0}H > 1.8~T_c$, the paramagnetic state would be lower in  energy . We shall refer to this field as $H_P$. $D(E_F)$ cancels from both sides of the expression leading to a certain degree of material independence. This  is referred to as the Pauli or the Clogston-Chandrasekhar limit in literature \cite{clogston1962upper, chandrasekhar1962note}. Upper critical field of elemental superconductors are far below this limit. There has been a number of efforts to find exotic superconductors (e.g FFLO, spin-triplet, topological ) in recent past which might be compatible with relatively high magnetic field. Several methods including metal intercalation between layers of a material \cite{hor2010superconductivity, asaba2017rotational}, applying high pressure, hard and soft tip contact \cite{hou2019superconductivity, aggarwal2016unconventional, hou2020two} and interface between two materials \cite{zhu2020interfacial,kononov2020flat} has been tried for achieving this. In this paper, we point out an emerging class of materials, composed of ${\rm Ga}$ and a heavy metal carbide, fabricated using FIB which are undoubtedly granular and highly disordered, but with  $H_{c2}$   significantly above the Pauli limit. \\

A metalorganic gas,  with a heavy metal (${\rm W}$, ${\rm Pt}$, ${\rm Nb}$) bonded to organic functional groups is injected close to the area of interest on a sample. Either an accurately directed electron beam or a ${\rm Ga^{+}}$ beam is used to decompose the gas, leaving behind a residue of ${\rm Ga}$, ${\rm C}$ and the heavy metal. The stoichiometry of the residue is not precise but its superconducting properties are robust \cite{sadki2004focused,dobrovolskiy2020ultra, cordoba2019long, porrati2019crystalline, porrati2017electrical, guillamon2008nanoscale,sun2013voltage, cordoba2013magnetic,sengupta2015superconducting, chakraborti2018coherent}. In our samples,  hexacarbonyl tungsten $\rm{(W(CO)_6)}$ was selectively decomposed by a scanning ${\rm Ga^{+}}$ beam  leaving an amorphous ${\rm Ga{\text -}W{\text -}C}$ residue at programmed locations forming the tracks  (Table \ref{tab:table1}). \\

\begin{table}
	\begin{center}
		\caption{The sample parameters. Resistivity was calculated using  $ \rho_n = \frac{L_y\times L_z\times R_N}{L_x} $; where, $ L_x $, $ L_y $, $ L_z $ are length, width and thickness of those samples respectively. The justification of treating this as 3D rectangular blocks comes from the fact that the mean free path and coherence lengths are both much smaller than the shortest of the three dimensions in all the samples. Orbital limiting field is calculated using the relation: $ H^{orb}_{c2}(0) = 0.7\left|T\dfrac{dH_{c2}}{dT}\right|_{T_c} $ .}
		\label{tab:table1}
		\begin{tabular}{c|c|c|c|c}
			
			\hline

			\multicolumn{1}{p{1cm}|}{\centering ID \\ (Width)\\nm}&
			\multicolumn{1}{|p{2cm}|}{\centering Deposition \\ parameter\\kV, pA}&
			\multicolumn{1}{|p{1cm}|}{\centering $T_c $ \\ ~\\$\rm {K} $}&
			\multicolumn{1}{|p{2cm}|}{\centering $ \rho_n $\\~  \\ $ \mu \Omega.$ cm}&
			\multicolumn{1}{|p{1cm}}{\centering $ H^{orb}_{c2}(0) $ \\~ \\ $ \rm{T} $}\\
			\hline
			
			A ($ 100 $ ) & $ 30 $, $ 52 $ & $ 5.20 $ & $ 65 $& $ 18.0 $\\
			B ($ 500 $ ) & $ 30 $, $ 52 $ &  $ 5.02 $ & $ 110 $& $ 12.4 $\\
			C ($ 1000 $ ) & $ 30 $, $ 52 $ &  $ 5.10 $ & $ 115 $&$ 13.0 $\\
			D01 ($ 440 $) & $ 30 $, $ 20 $ & $ 4.50 $ & $ 125 $&$ 16.8 $\\
			D02 ($ 280 $) & $ 30 $, $ 20 $ &$  5.02 $ & $ 110 $&$ 14.0 $\\
			\hline
		\end{tabular}
	\end{center}
\end{table}

\begin{figure}
	\begin{center}
		\includegraphics[width=0.5\textwidth]{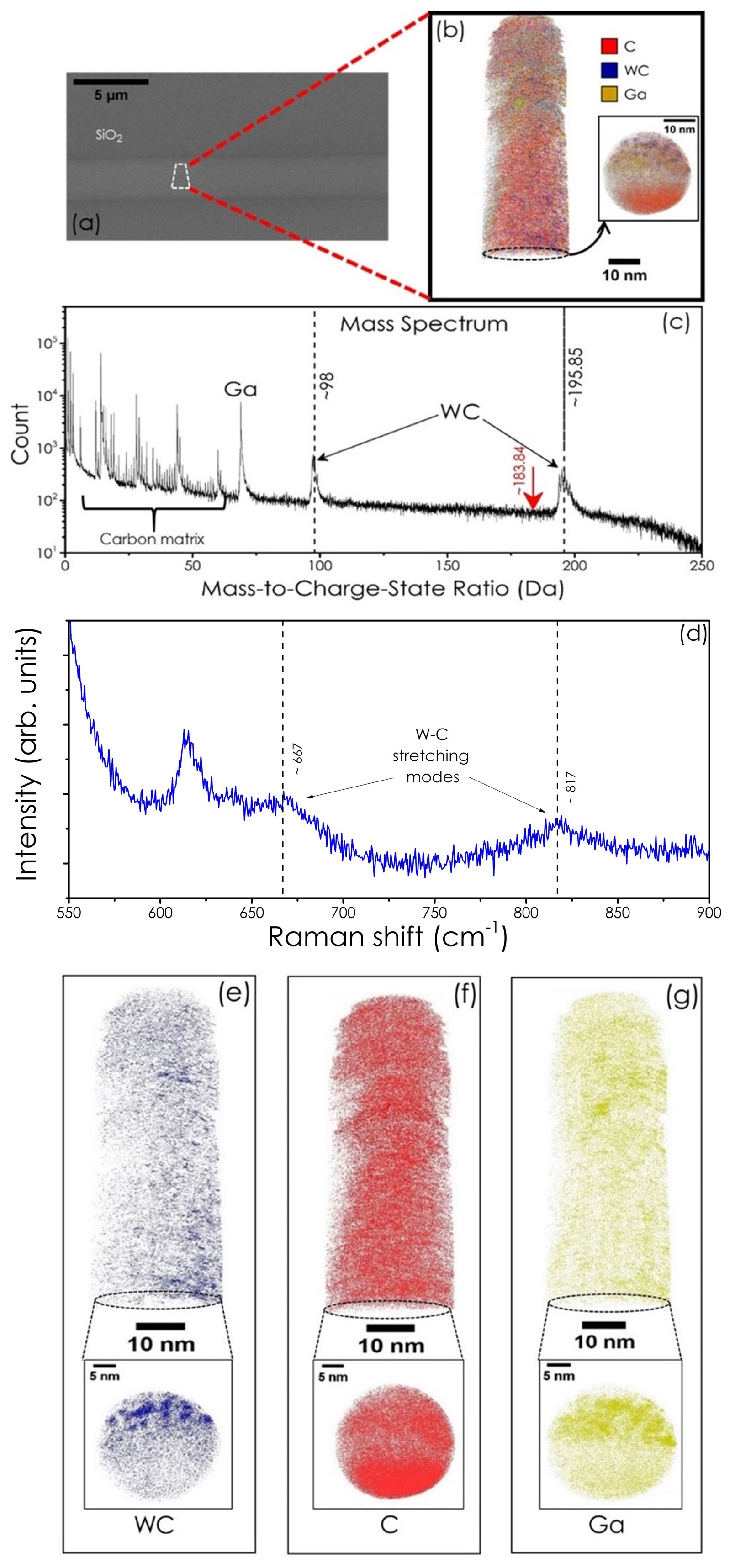}

	\end{center}
	\caption{\small{(a)Nanowire sample with region of interest (ROI) marked in it. (b) 3D reconstruction of APT data of ROI. (c) Mass spectrum showing the presence of singly and doubly ${\rm WC^{+}}$ and ${\rm WC^{+2}}$ but no free ${\rm W}$. (d) Raman spectrum obtained from the sample.  (e), (f) and (g) The distributions of the major components ${\rm WC}$, ${\rm C}$, and ${\rm Ga}$ in the composite. Cross sectional distribution  is shown in the bottom panel of each image. As can be seen in figure- (f), most of the carbon is precipitated near the substrate. So the major constituents of the upper half of the sample is ${\rm Ga}$ and ${\rm WC}$. Our DFT calculation of ${\rm Ga}$ decorated ${\rm WC}$ surface is guided by this observed distribution.
	}}\label{apt1}
\end{figure}
To understand the microstructure of the residue, we carried out Atomic Probe Tomogrpahy (APT) analysis of the tracks. In APT, the sample is prepared in the form of a needle, and atom by atom \enquote{evaporation} from the sample is done  utilizing a strong electric field \cite{blavette1993atom}. Ions are evaporated from the  apex and projected into a position sensitive single ion detector \cite{bas1995general}. From the positions and time-of-flight data, 3D reconstruction of the analyzed volume is done. It has a lateral resolution of $ 0.3-0.5 $ nm and a depth resolution of $ 0.1-0.3 $ nm. The constituents of the  sample can be identified from this mass spectrum.  The measurements were done using CAMECA local electrode atom probe (LEAP 5000 XR). Tip evaporation was carried out in laser mode with pulse energy $ 30~ \rm {pJ}$  and frequency  and $ 200~\rm{kHz} $  at $ 40~\rm{K}$.
Certain clear conclusions can be drawn from the data:
\begin{enumerate}
	\item The key observation is that tungsten  is in the form of its carbide (WC). The two peaks (figure \ref{apt1}(c)) correspond to singly and doubly ionised species with mass $ 196 $ amu. There is no W ($ \sim 184 $ amu) in free form. It is important to emphasize that doubly/triply ionised ions cannot give rise to peaks at high amu. The absence of peaks corresponding to ${\rm W^{+}}$, ${\rm W^{+2}}$ or ${\rm W^{+3}}$ and the presence of peaks at $98$ and $196$ amu unambiguously show the  formation of the carbide and absence of the free metal.

\item Raman spectra (Fig \ref{apt1}d) also shows some evidence of the ${\rm WC}$ stretch modes  at $667{\rm cm^{-1}}$ and $817 {\rm cm^{-1}}$ identified in ref \cite{yang2008effect,Porrati_2010} 
 		
\item The abundance of ${\rm W}$, ${\rm C}$ and ${\rm Ga}$ calculated from APT  agrees  well with  energy dispersive X-ray spectroscopy (EDS) carried out separately  at various locations of the samples. Atomic composition of all the  samples are similar : C $(\approx 52\%)$, W $(\approx 36\%)$ and Ga $(\approx 12\%)$.
	\item  There is noticeable precipitation (figure \ref{apt1}(f)) of  carbon towards the bottom of the nanowires. This may be of some significance as thin polarizable underlayers of materials like ${\rm Ge}$ (for example) have been shown to increase the $T_c$ of strongly disordered quench condensed films of metals. Such polarizable dielectrics may modify the electron-electron interactions in the granular thin films \cite{haviland1989onset,gupta2001possible, gupta2002critical, lita2005tuning, sambandamurthy2001effect, witanachchi1989effect}.
\end{enumerate}

\begin{figure}
	\begin{center}
		\includegraphics[width=0.5\textwidth]{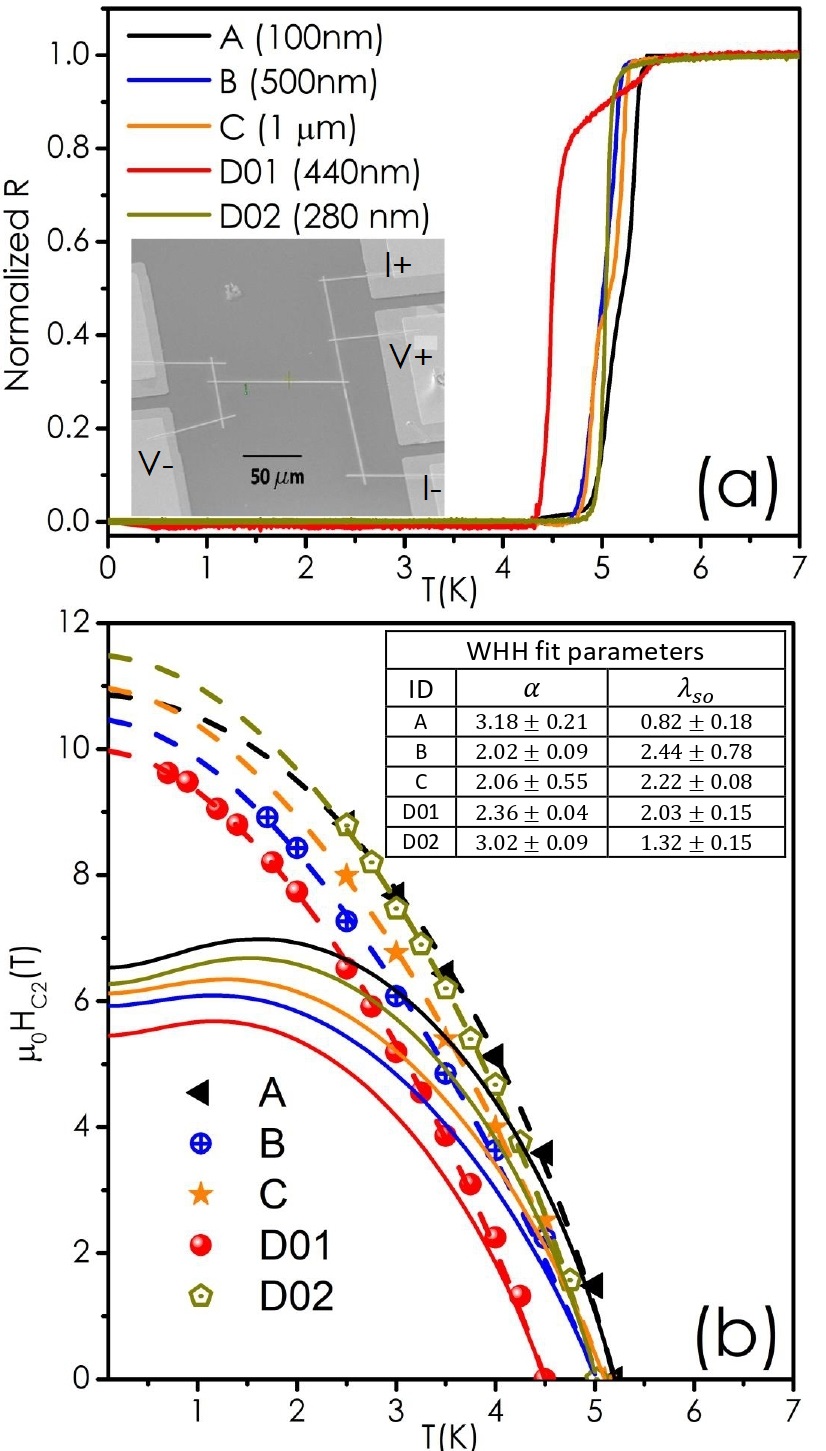}
	\end{center}
	\caption{\small{(a)The superconducting transition of the five samples studied, the inset shows an electron micrograph of a typical sample, configured for 4-terminal measurements. (b) The $H_{c2}(T)$ data for all the samples and the WHH fits (dotted lines). The table (inset) shows the fit parameters $\alpha$ and $\lambda_{so}$ along with fit errors. The solid lines are obtained by setting $\lambda_{so}=0$, for the same value of $\alpha$ showing how the existence of the spin-orbit coupling pushes up the critical field. The magnetic field was applied along the direction of current flow.}}
	\label{fig1_WHH}
\end{figure}

Fig \ref{fig1_WHH}(a)  shows the typical configuration of the contacts (inset) used for four-terminal transport measurements and a sharp zero field resistive transition of all the samples. The $H_{c2}(T)$ curves (Fig \ref{fig1_WHH}(b)) were obtained by fixing the temperature and sweeping the magnetic field. The point where the resistance reached half the normal state value was taken as $H_{c2}$. The maximum field that could be applied was $10 {\rm T}$ and the lowest temperature was $T\approx 260 {\rm mK}$. The $H_{c2}(0)$ values were obtained by extrapolation of the fitted curve (Fig \ref{fig1_WHH}(b)). The $T = 0$ value is clearly higher than the Clogston\textendash Chandrasekhar limit. None of the two well studied allotropic forms of ${\rm W}$, namely $\alpha$ and $ \beta $-tungsten can explain this behaviour.\cite{gibson1964superconductivity, basavaiah1968superconductivity,johnson1966superconductivity, lau2020concomitance}
 The observed upper critical field arises as a combined effect of the orbital and spin paramagnetism related pair breaking. The manner in which these two parts combine may be described by the Werthamer-Hohenberg-Helfand (WHH) theory \cite{maki1964pauli, werthamer1966temperature, helfand1966temperature}. $ \mu_0{H_{c2}}(T)$ curves depend on two important dimensionless parameters, the Maki parameter $\alpha = \sqrt{2}H_{c2}^{orb}/H_P$  and $\lambda_{so}$ the spin orbit scattering strength. The limiting orbital field, $H_{c2}^{orb}$  may be calculated from experimental data as $H_{c2}^{orb} (T=0)\approx 0.7{\left|T\dfrac{dH_{c2}}{dT} \right|_{T_c}}$. The applicability of WHH theory to amorphous superconductors is well established \cite{poon1983analysis, carter1981enhanced, hofer2019superconductivity}. Later on we shall show that the application of the WHH theory is justified for this system. However it is also important to point out that there are known examples where the WHH framework doesn't work \cite{gantmakher1996positive, lu2015evidence}. 
Figure \ref{fig1_WHH}(b) shows the behaviour of five superconducting samples listed in table \ref{tab:table1} along with the fit-parameters. 
These fits are described by
\begin{eqnarray}
\begin{aligned}
\ln\left(\frac{1}{t}\right)=\sum_{\nu=-\infty}^{\infty}\left(\frac{1}{|{2\nu+1}|}-\left[|{2\nu+1}|+\frac{b_c}{t} \right. \right.+ \\
\left.\left.\frac{(\alpha {b_c}/t)^2}{|{2\nu+1}|+(b_c+\lambda_{so})/t}\right]^{-1}\right)
\end{aligned}
\label{WHHeqn}
\end{eqnarray}
where $t= T/T_c$,  $\alpha=  \frac{\sqrt{2}H_{c2}^{orb}}{H_{P}}$ is the Maki parameter, the dimensionless magnetic field $b_c = {\mu_0}{h_c} = \dfrac{e{\hbar}}{2m\pi\alpha}\dfrac{{\mu_0}H_{c2}}{{k_B}{T_c}}$ and $\lambda_{so}$ is the spin-orbit scattering constant \cite{maki1964pauli}.  Figure \ref{fig1_WHH}(b) shows the values of  the spin-orbit scattering and Maki parameter  obtained from these fittings. Critical field enhancement of $ 1.15 $ to $ 1.27 $ times $H_P$  has been observed in our samples. We find an overall trend of $\alpha$ decreasing with increasing thickness and $\lambda_{so}$ increasing with increasing thickness. The  value of $\alpha$ ($=2{\text -}3$) is more than the $ 1.8 $ in all the films, suggesting \cite{matsuda2007fulde} that this system can potentially host an FFLO phase. The values of $\lambda_{so}$ points to the presence of moderate spin-orbit scattering in this system that  can enhance the $H_{c2}$ significantly by suppressing the  spin paramagnetism. This has been observed in dirty layered superconductors consisting of transition metal dichalcogenide intercalated with organic molecules, with strong SOC \cite{gamble1970superconductivity,klemm1975theory}.

We now validate the  three key requirements for the applicability of WHH analysis :  an isotropic Fermi surface, weak electron-phonon coupling ($ \lambda_{ep} $ ) and the rate of spin-flip scattering ($ 1/\tau_{so} $) being lower than the non-spin-flip ($ 1/\tau_{tr} $)scattering. In homogeneously disordered samples like these fermi surface is almost certainly isotropic.
For all our samples, the transport scattering time $ \tau_{tr}= l/v_F < 10^{-16}~\rm sec $. $ \tau_{so} $ is obtained from the WHH fit using $ \tau_{so}=\dfrac{\hbar}{3\pi k_BT_c\lambda_{so}}\approx 10^{-13}~\rm sec $.  It is clear that $1/ \tau_{tr} \gg 1/\tau_{so}$.
The electron-phonon coupling strength $ \lambda_{ep} $ can be estimated using the McMillan equation \cite{mcmillan1968transition}
\begin{equation}
\lambda_{ep} = \frac{1.04+\mu^*ln(\frac{\Theta_D}{1.45T_c})}{(1-0.62\mu^*)ln(\frac{\Theta_D}{1.45T_c})-1.04}
\label{macmil}
\end{equation}
The  Coulomb pseudopotential, $ \mu^*$  for metal films is typically $\approx  0.1 - 0.15 $. Debye temperatures of crystalline tungsten and its alloys are typically $\Theta_D \approx 400 {\rm K}$ \cite{liu2019thermodynamic}. For device D01 we obtain $ \lambda_{ep} \sim 0.5 $ consistent with the weak coupling assumptions \cite{foner1981superconductor, toyota1984upper}.
\\

\begin{figure}
	\centering
	\includegraphics[width=0.5\textwidth]{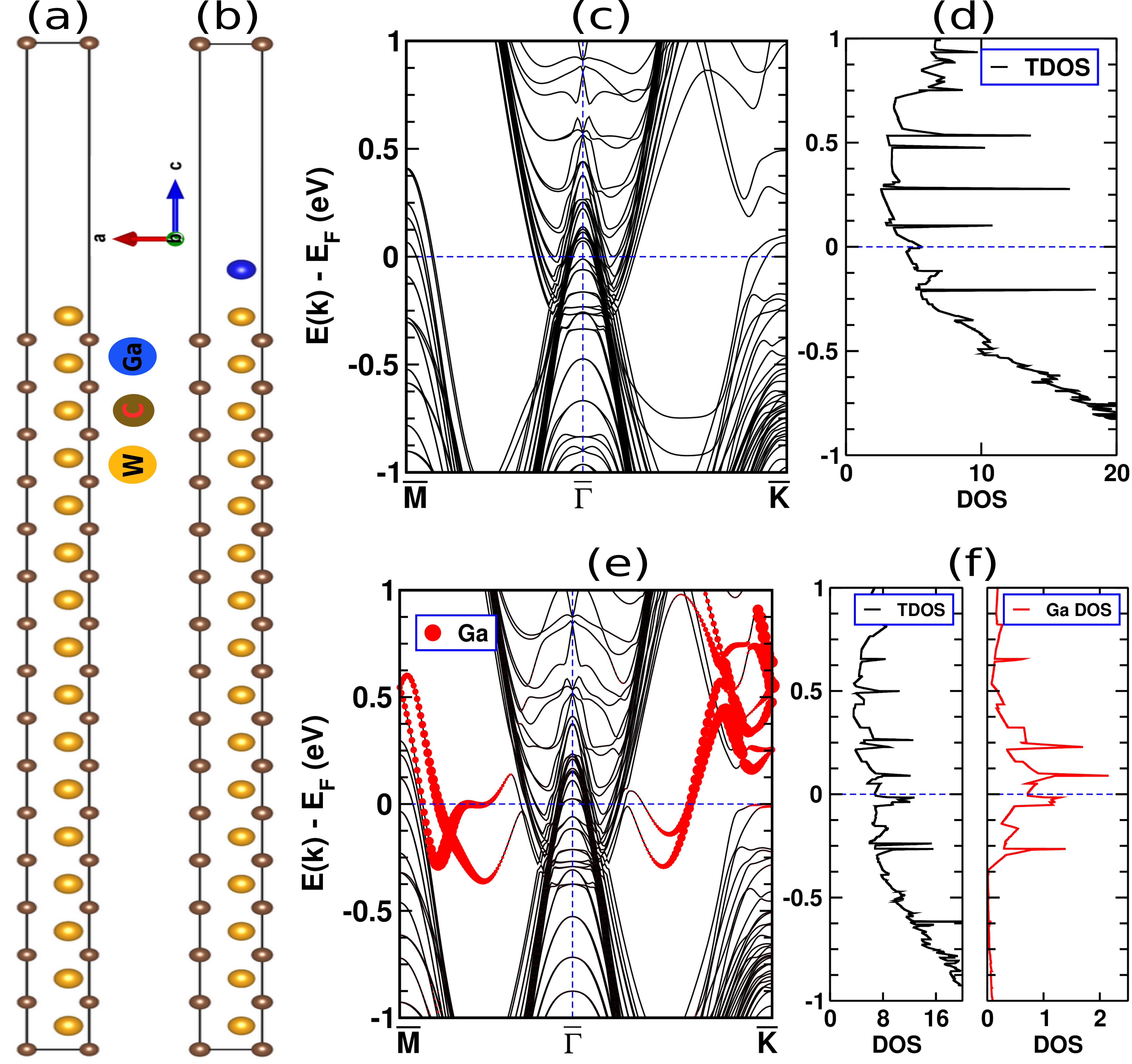}
	\caption{(Color online)DFT results using plane wave basis set using the projector augmented wave (PAW) \cite{blochl1994projector} method  with an energy cutoff of 550 eV. Exchange and correlation was incorporated using  generalized-gradient approximation by Perdew, Burke, and Ernzerhof \cite{kresse1999ultrasoft}. Total energy(force) convergence criterion was set to 10$^{-5}$ eV (0.01 eV/\r{A}). (001) Surface slabs for (a) WC and (b) WC with Ga adsorbant. (c) and (d) shows the band structure and total density of states(TDOS) for WC without Ga, while (e) and (f) represent the same with Ga adsorbant. Total density of states (TDOS) of the slab and partial density of states of Ga atom are shown by black and red colored line in (f), respectively. Density of states are plotted in the unit of states/eV/Cell. Red dots in (e) represent the Ga atom contribution. Relative size of the dots represents the projected contribution of Ga to the total DOS.  }
	\label{surfaceslab}
\end{figure}

To understand the possible role of Ga, we have performed density functional electronic structure calculations for Ga adsorbed on WC system, using the Vienna ab initio simulation package(VASP) \cite{kresse1993ab,kresse1999ultrasoft}.   WC crystallizes in the space group P$\bar{6}$m2, with the chosen lattice constants $a=b=2.198$ \r{A}, and $c=2.846$ \r{A} \cite{ma2018three}. A  (001) surface slab with 16 unit cell was constructed with ${\rm W}$ termination on the top and ${\rm C}$ on the bottom surface respectively. A vacuum of $\sim$15 \r{A} was added to nullify the interaction between the top and bottom surfaces. We added a single ${\rm Ga}$ adsorbant on the top surface of the slab (figure~\ref{surfaceslab}(a,b)). ${\rm Ga}$  on ${\rm W}$ was found to be energetically more favourable than on ${\rm C}$ sites.  The Brillouin zone (BZ) integration of the slab was performed using a $11\times 11\times 1$ $\Gamma$-centered $k$-mesh. We  also performed the ionic relaxation to fully optimize the slab geometry.
Figure~\ref{surfaceslab}(c, e) shows the band structure for the surface slabs of WC without and with Ga adsorbant ( figure \ref{surfaceslab}(a) and (b)) respectively. Both the surfaces show a large band density near  $E_F$. In figure~\ref{surfaceslab}(e), there are additional bands that arise from the ${\rm Ga}$ adsorbant on the surface of the slab (shown by red-colored dots). DOS (in states/eV/Cell), of bare WC and Ga adsorbed WC are shown in figure~\ref{surfaceslab}d\&f respectively. Figure~\ref{surfaceslab}f clearly shows significant contribution of ${\rm Ga}$ to DOS near $E_F$, which mainly arises from the relative flat nature of the Ga-projected bands in figure~\ref{surfaceslab}e. Such Ga-mediated bands can be the origin for the observed $ T_c $ enhancement \cite{sengupta2015superconducting}, from nearly $ 2~\rm K $ to $ 5~\rm K$, in Ga beam prepared composite as compared to that of electron beam prepared one.\\

Another point that needs to be addressed is to what extent granularity can enhance the critical field in these samples. In such samples the superconducting order parameter tends to stay nearly constant within the small grains as the variation over such small length scales will come with a significant free energy cost. The nearby grains are then Josephson coupled leading to a global superconducting state. However as shown by Deutscher et al \cite{PhysRevB.22.4264}, this is accompanied by a change in the curvature of $H_{c2}(T)$ curves at some intermediate temperature as the disorder enhancement happens. We see no such (inflection point) feature in any of our samples with the thickness varying from $100~{\rm nm}$  to $1~\mu{\rm m}$. Secondly quench-condensed films\cite{gupta2001possible, gupta2002critical, sambandamurthy2001effect},
 electrodeposited nanowires \cite{Tian_Bi_Nanowire} in which superconductivity has been shown to occur due to formation of metastable phases of ${\rm Bi}$, ${\rm Sn}$ etc, are
known to be unstable against thermal cycling. The metastable phases relax back to the non-superconducting stable phase. We have observed no change in the characteristics of the ${\rm WC{\text{-}}Ga}$ samples on repeated thermal cycles to room temperature. The high critical fields of these does not appear to be a granularity driven effect. \\

Finally we point out that recently several independent and interesting  results have been obtained about superconductivity in heavy metal carbides. High critical fields and supersonic vortex velocities have been reported in ${\rm Nb}$ tracks made using FIB \cite{dobrovolskiy2020ultra}. Our observations of vortex velocities in WC system are also very similar. $H_{c2}$ above Pauli limit has been reported in Rhodium-Carbide \cite{acsmaterialsau.1c00011}.  At the same time  recent band structure calculation supported by ARPES data identified non-trivial topological aspects of surface states of WC \cite{ma2018three}. The interface of single-crystal WC and metal have shown superconductivity till considerably high magnetic fields \cite{zhu2020interfacial, hou2019superconductivity}. The large interfacial area of the nanocrystals with the Ga-C composite may be fulfilling similar conditions in these samples. It has been predicted recently \cite{sim2019triplet,lin2020chiral} that materials having triple-band crossing, like in WC, can uniquely stabilize spin-triplet superconductivity. FIB patterned carbide superconductor tracks could thus emerge as an important component for junctions and interconnects in superconducting electronics and a host for novel superconducting states.

\section{Acknowledgements}  
We are thankful for several discussions with  B. Karmakar, A. Pal, H. Suderow, A. Taraphder, S. Mahapatra. We acknowledge the national facility for APT at IIT Madras, India and thank A. Kumbhar, R. Gupta and G. Goyal for helping with APT procedures. We acknowledge  Raith GmbH, Bochum and MEMS department IIT Bombay for the FIB samples. We  acknowledge funding from Department of Science and Technology, Government of India under project: SR/S2/CMP-71/2012, DST-FIST and the central facilities of IIT Bombay.

\medskip

\bibliographystyle{apsrev4-2}
\bibliography{ref}
\end{document}